\documentclass[12pt,a4paper]{article}
\usepackage{amsfonts,latexsym}
\usepackage{amsmath,amssymb}
\usepackage{graphicx,color}

\renewcommand{\thefootnote}{\fnsymbol{footnote}}

\begin{document}

\vspace{12mm}
\begin{center}
{{{\Large {\bf Quantum parameter-mass induced   scalarization of qOS-black hole in the Einstein-Gauss-Bonnet-scalar theory }}}}\\[10mm]
{Yun Soo Myung\footnote{e-mail address: ysmyung@inje.ac.kr}}\\[8mm]

{Center for Quantum Spacetime, Sogang University, Seoul 04107, Republic of  Korea\\[0pt] }

\vspace{12mm}

\end{center}
 \begin{abstract}
    \noindent We obtain quantum parameter  ($\alpha$)-mass ($M$) induced spontaneous  scalarization of quantum Oppenheimer-Snyder (qOS)-black hole in the Einstein-Gauss-Bonnet-scalar theory with the unknown qOS action.
     We derive Smarr formula which describes   a correct thermodynamics  for the bald qOS-black hole.
    It is turned out that two Davies points of heat capacity are identified with two critical onset mass and quantum parameter for spontaneous scalarization.
    However, we do not obtain such connections from the Einstein-Gauss-Bonnet-scalar theory with the nonlinear electrodynamics action.
      Furthermore, the shadow radius analysis of qOS-black hole is performed to distinguish quantum parameter  from  mass  by comparing them  with the EHT observation.
      There is no constraints on the quantum parameter, but new constraints are found for the mass.
      This work  is considered as the first example to show a close connection between thermodynamics of bald black hole and spontaneous scalarization.
\end{abstract}

\vspace{1.5cm}

\hspace{11.5cm}{Typeset Using \LaTeX}
\newpage
\renewcommand{\thefootnote}{\arabic{footnote}}
\setcounter{footnote}{0}

\vspace{2mm}


\section{Introduction}
\label{sec: Intro}
The no-hair theorem states that a black hole can be completely characterized by only three externally observable classical parameters: mass, electric charge, and angular momentum in general relativity
\cite{Carter:1971zc,Ruffini:1971bza}. If a scalar field is minimally coupled to  gravitational and electromagnetic  fields, there is no scalar hair~\cite{Herdeiro:2015waa}.
However, its evasion  occurred in the context of scalar-tensor theories possessing the nonminimal scalar coupling to either Gauss-Bonnet (GB) term~\cite{Doneva:2017bvd,Silva:2017uqg,Antoniou:2017acq}, or to Maxwell term~\cite{Herdeiro:2018wub,Myung:2018vug}, where the former is called GB$^+$ scalarization arisen  from tachyonic instability.

Some authors have  investigated the  spin-induced (GB$^-$) scalarization
of Kerr black holes in the Einstein-Gauss-Bonnet-scalar (EGBS)
theory with a negative  coupling parameter~\cite{Cunha:2019dwb,Collodel:2019kkx}.
Later on, an $a$-bound of $a \ge a_c=0.5M$ was found to represent the onset of scalarization for
Kerr black holes~\cite{Dima:2020yac}.
This critical rotation parameter $a_c$ was  computed  not only analytically~\cite{Hod:2020jjy} but also
numerically~\cite{Zhang:2020pko,Doneva:2020nbb}. This implies that the low rotation of $a<a_c$ suppresses spontaneous scalarization.
To this direction, the spin-induced scalarized black holes were numerically
constructed for high rotation and negative coupling parameter in the EGBS theory~\cite{Herdeiro:2020wei,Berti:2020kgk}.
Furthermore, the spin-induced scalarization of Kerr-Newman black holes was found for $a\ge a_c=0.672M$ with $Q=0.4$~\cite{Lai:2022spn} and the spin-charge induced scalarization of Kerr-Newman black holes was studied in Einstein-Maxwell-scalar theory~\cite{Lai:2022ppn}.

On the other hand,  the quantum Oppenheimer-Snyder (qOS)-black hole was recently  discovered from studying the qOS gravitational collapse within the loop quantum cosmology~\cite{Lewandowski:2022zce}.
Therefore, we do not know  its action $\mathcal{L}_{\rm qOS}$ to give the qOS-black hole as a direct solution.
In this model, the singularity is replaced by a transition region that includes an inner horizon.
Even though various aspects of this model were explored, we wish to stress on  thermodynamics~\cite{Dong:2024hod}, shadows~\cite{Ye:2023qks}, and its scalarization~\cite{Chen:2025wze}.
Very recently, it is desirable to note that its energy-momentum tensor was known and its action was proposed  by the nonlinear electrodynamics action~\cite{Mazharimousavi:2025lld}.

In this study, we wish to  revisit thermodynamics of qOS-black holes described by mass $M$ and quantum parameter $\alpha$, and GB$^-$ scalarization in the EGBS theory with the  unknown  qOS action.
We perform  thermodynamic aspects   by correcting thermodynamic quantities, checking  the first law thermodynamics and the Smarr formula, and studying a phase transition of heat capacity at Davies point.
The shadow radius analysis of qOS-black hole will be included  to distinguish quantum parameter $\alpha$  from  mass $M$   by comparing them  with the EHT observation.
Here, $\alpha$-$M$ induced scalarizations are  found  in the EGBS theory with the unknown qOS action: $\alpha_c(=1.2835)<\alpha\le \alpha_e(=1.6875)$  with $M=1$ and $M_{\rm rem}(=0.7698)\le M <M_c(=0.8827)$  with $\alpha=1$ for conditions of spontaneous scalarization.
Importantly, we find that two Davies points ($\alpha_D,~M_D$) of heat capacity are identified with two critical onset quantum parameter ($\alpha_c$) and mass ($M_c$)  for spontaneous scalarization.
However, we do not obtain such connections from the Einstein-Gauss-Bonnet-scalar theory with the nonlinear electrodynamics action proposed by~\cite{Mazharimousavi:2025lld}.
We would like to stress that this work  is considered as the first model to show a close connection between thermodynamics of bald black hole and spontaneous scalarization.


\section{qOS-black hole  solution}
We introduce the Einstein-Gauss-Bonnet-scalar theory with the unknown qOS action~\cite{Chen:2025wze}  as
\begin{equation}
\mathcal{L}_{\rm EGBSq}=\frac{1}{16 \pi}\Big[ R-2\partial_\mu \phi \partial^\mu \phi+\lambda f(\phi) {\cal R}^2_{\rm GB}+{\cal L}_{\rm qOS}\Big],\label{Action1}
\end{equation}
where $\phi$ is the scalar field,  a coupling function $f(\phi)$, $\lambda$ is the GB coupling constant having length dimension two, and ${\cal R}^2_{\rm GB}$ is the GB term defined by
\begin{equation}
{\cal R}^2_{\rm GB}=R^2-4R_{\mu\nu}R^{\mu\nu}+R_{\mu\nu\rho\sigma}R^{\mu\nu\rho\sigma}.\label{Action2}
\end{equation}
Examples for coupling function include quadratic coupling $f(\phi)\sim \phi^2$~\cite{Silva:2017uqg} and exponential coupling $f(\phi)\sim (1-e^{-6\phi^2})$~\cite{Doneva:2017bvd}.
Other examples appeared in Ref.~\cite{Antoniou:2017acq}.

From the action (\ref{Action1}), we derive  the Einstein  equation
\begin{eqnarray}
 G_{\mu\nu}=2\partial _\mu \phi\partial _\nu \phi -(\partial \phi)^2g_{\mu\nu}+\Gamma_{\mu\nu}+T^{\rm qOS}_{\mu\nu}, \label{equa1}
\end{eqnarray}
where $G_{\mu\nu}=R_{\mu\nu}-(R/2)g_{\mu\nu}$ is  the Einstein tensor and  $\Gamma_{\mu\nu}$   is given by
\begin{eqnarray}
\Gamma_{\mu\nu}&=&2R\nabla_{(\mu} \Psi_{\nu)}+4\nabla^\alpha \Psi_\alpha G_{\mu\nu}- 8R_{(\mu|\alpha|}\nabla^\alpha \Psi_{\nu)} \nonumber \\
&+&4 R^{\alpha\beta}\nabla_\alpha\Psi_\beta g_{\mu\nu}
-4R^{\beta}_{~\mu\alpha\nu}\nabla^\alpha\Psi_\beta  \label{equa2}
\end{eqnarray}
with
\begin{equation}
\Psi_{\mu}=\lambda \frac{df(\phi)}{d\phi} \partial_\mu \phi=\lambda f'(\phi)\partial_\mu \phi.
\end{equation}
Here, even though $\mathcal{L}_{\rm qOS}$ is not known~\cite{Lewandowski:2022zce}, its energy-momentum tensor should take the form to find $g(r)$ in Eq.(\ref{g-sol})~\cite{Mazharimousavi:2025lld}
\begin{equation}\label{q-em}
T^{\rm qOS, \nu}_{\mu}=\frac{3\alpha M^2}{r^6}{\rm diag}[-1,-1,2,2].
\end{equation}
Its suggesting  action may be  given by the nonlinear electrodynamics action
\begin{equation} \label{NED}
\mathcal{L}_{\rm NED}=\frac{1}{16\pi}\Big[2\xi(-\mathcal{F})^{\frac{3}{2}}\Big]
\end{equation}
with  $\mathcal{F}=F^{\mu\nu}F_{\mu\nu}=2P^2/r^4$ for a magnetic charge configuration with $F_{\theta\varphi}=P \sin \theta$.
In this case, choosing  $\alpha=2^{3/2}\xi P/3$ with  $P=M$  may lead to Eq.(\ref{q-em}) because its energy-momentum tensor is determined  by
\begin{equation}
T^{\rm NED,\nu}_\mu=\frac{2^{3/2}\xi P^3}{r^6}{\rm diag}[-1,-1,2,2].
\end{equation}
However, selecting $P=M$ is not allowed for non-extremal black holes. We will treat this case in Section 6 separately.

On the other hand, the scalar field equation takes the form
\begin{equation}
\square \phi +\frac{\lambda}{4}f'(\phi) {\cal R}^2_{\rm GB}=0 \label{s-equa}.
\end{equation}
Solving $G_{\mu\nu}=T^{\rm qOS}_{\mu\nu}$ with $\phi=0$ and $f'(\phi)|_{\phi=0}=0$, one finds the qOS-black hole  solution
\begin{equation} \label{ansatz}
ds^2_{\rm qOS}= \bar{g}_{\mu\nu}dx^\mu dx^\nu=-g(r)dt^2+\frac{dr^2}{g(r)}+r^2d\Omega^2_2
\end{equation}
where  the metric function is given by~\cite{Lewandowski:2022zce}
\begin{equation}
g(r)=1-\frac{2M}{r}+\frac{\alpha M^2}{r^4} \label{g-sol}
\end{equation}
with the quantum parameter $\alpha$ having length dimension two.
We note  that (\ref{ansatz})  indicates  the qOS-black hole solution without scalar hair.

\section{Smarr formula and thermodynamic analysis for qOS-black hole}
In this section, we wish to study the thermodynamics of qOS-black hole with Smarr formula  in the grand canonical ensemble. In this ensemble, the entropy $S$ and quantum parameter $\alpha$ change with the surrounding heat bath, while temperature and chemical potential are fixed.
\begin{figure}
\centering
\mbox{
(a)
\includegraphics[angle =0,scale=0.4]{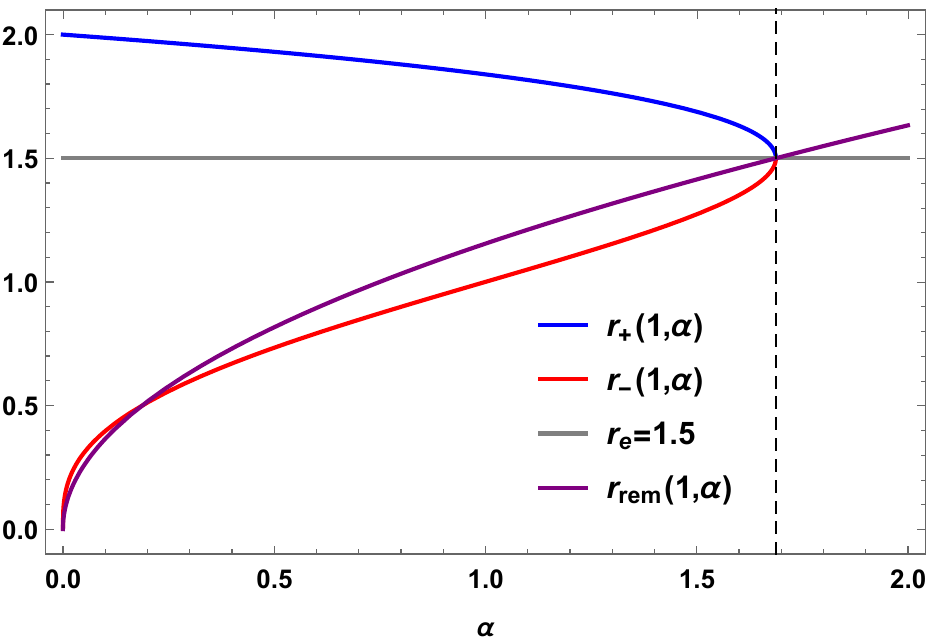}
(b)
\includegraphics[angle =0,scale=0.4]{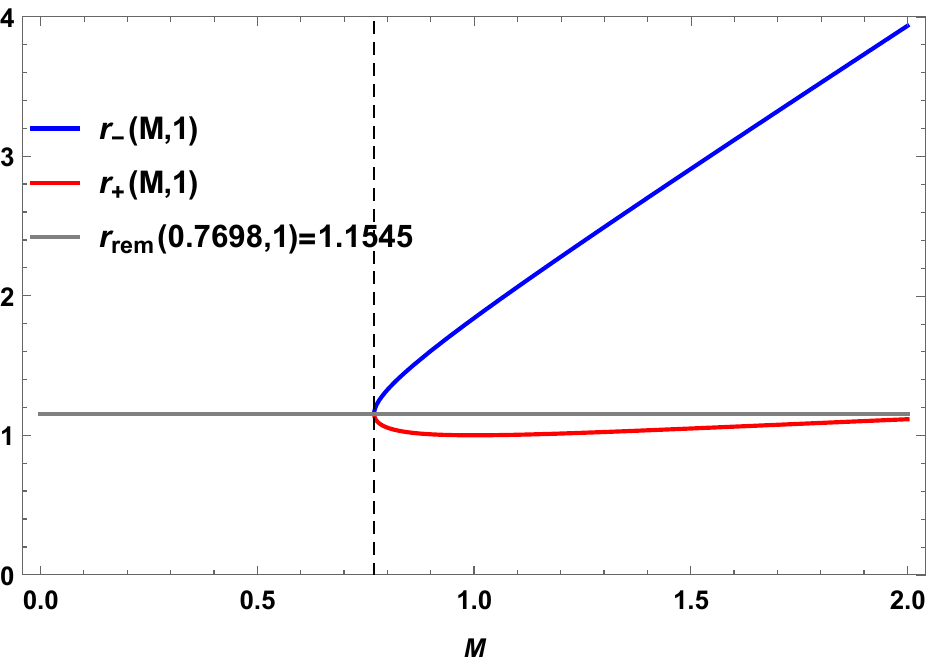}
}
\caption{ (a) Two outer/inner horizons $r_\pm(M=1,\alpha)$ are  functions of $\alpha\in[0,1.6875]$, indicating the upper bound as an extremal point. The remnant radius $r_{\rm rem}(1,\alpha)=2\sqrt{\alpha}/\sqrt{3}>r_-(1,\alpha)$ as function of $\alpha$ represents the  transition region including the inner horizon $r_-(1,\alpha)$.  (b) Two  horizons  $r_{\pm}(M,\alpha=1)$ are   function of   $M\in[0.7698,2]$, showing the lower bound for the mass of black hole. Here, $r_{\rm rem}(0.7698,1)=1.1545>r_-(M,1)$ represents a constant.}
\label{plot_modes}
\end{figure}
From $g(r)=0$, one finds two real solutions and two complex solutions
\begin{eqnarray}
&&r_i(M,\alpha),~{\rm for}~i=4,3,2,1, \label{f-roots} 
\end{eqnarray}
where $r_4(M,\alpha)\to r_+(M,\alpha)$ and $r_3(M,\alpha)\to r_-(M,\alpha)$ and $r_{2/1}$ become complex solutions. As is shown in Fig. 1, there is an upper bound on $\alpha=1.6875$ as an extremal point for $M=1$, while there is a lower bound for the mass of  black hole (remnant mass $M_{\rm rem}$=0.7698 for $\alpha=1$).

We find that black hole mass $m(M,\alpha)$ obtained from $g(r_+)=0$  after replacing $M$ with $m$, area-law entropy $S=\pi r_+^2$, the Hawking temperature defined by $T=\frac{\partial m}{\partial S}$,  heat capacity $C=\frac{\partial m}{\partial r_+}(\frac{\partial T}{\partial r_+})^{-1}$, chemical potential $W_{\alpha}=\frac{\partial m}{\partial \alpha}$, and Gibbs free energy $G=m-TS-W_{\alpha} \alpha$ as
\begin{eqnarray}
m(M,\alpha)&=&\frac{r_+^3(M,\alpha)-r_+^2\sqrt{r_+^2-\alpha}}{\alpha}\rightarrow_{r_+\to \sqrt{ S/\pi}} m(S,\alpha), \label{mass}\\
T(M,\alpha)&=& \frac{2\alpha -3r_+^2(M,\alpha)+3r_+ \sqrt{r_+^2-\alpha}}{ 2 \pi \alpha \sqrt{r_+^2-\alpha}} \rightarrow_{r_+\to \sqrt{ S/\pi}} T(S,\alpha),   \label{tem1} \\
 C(M,\alpha)&=&-\frac{2\pi r_+(M,\alpha)(r_+^2-\alpha)[2\alpha -3r_+^2+3r_+\sqrt{r_+^2-\alpha}]}{3(\alpha-r_+^2)\sqrt{r_+^2-\alpha}+3r_+^3-4 \alpha r_+},                         \label{heat1} \\
 W_{\alpha}(M,\alpha)&=& -\frac{ r_+^2(M,\alpha)[\alpha-2r_+^2+2r_+ \sqrt{r_+^2-\alpha}]}{2\alpha^2\sqrt{r_+^2-\alpha}} \rightarrow_{r_+\to \sqrt{ S/\pi}} W_\alpha(S,\alpha),                           \label{chem1l} \\
 G(M,\alpha)&=&\frac{r_+^2(M,\alpha)[r_+- \sqrt{r_+^2-\alpha}]}{2}.                        \label{Gibbs1}
\end{eqnarray}
Here, one finds that $m(S,\alpha),~T(S,\alpha), W_\alpha(S,\alpha)$ are found by replacing $r_+(M,\alpha)$ by $\sqrt{S/\pi}$.
From observing Eqs.(\ref{tem1}) and (\ref{heat1}), $T(M,\alpha)$ and $C(M,\alpha)$ are zero when their numerators are zero, showing extremal points,  while $C(M,\alpha)$ blows up when its denominator is zero, leading to the Davies point for a phase transition. In addition, we wish to define the heat capacity to match the Davies point with the critical onset parameter as
\begin{equation}
C(M,\alpha)\equiv \frac{NC(M,\alpha)}{DC(M,\alpha)}, \label{h-nd}
\end{equation}
whose Davies point can be obtained from solving $DC(M,\alpha)=0$.
We note that $T(M,\alpha)$ is not equal to  the surface gravity defined by~\cite{Dong:2024hod}
 \begin{equation}
 T_\kappa(M,\alpha)\equiv \frac{g'(r)|_{r\to r_+}}{4\pi}=\frac{2\alpha -3r_+^2(M,\alpha)+3r_+ \sqrt{r_+^2-\alpha}}{ 2 \pi \alpha r_+}
\end{equation}
whose denominator differs from that of $T(M,\alpha)$ in Eq.(\ref{tem1}) and thus, its heat capacity is different from $C(M,\alpha)$ in Eq.(\ref{heat1}). Importantly, if one uses $T_\kappa(M,\alpha)$, the following first law and Smarr formula are not satisfied.

At this stage, we  wish to check that  the first law of thermodynamics  is satisfied as~\cite{Myung:2025afs}
\begin{equation}
dm=T d S+W_{\alpha} d\alpha. \label{first-law}
\end{equation}
Also, the Smarr formula takes the form
\begin{equation} \label{smarr}
    m=2TS+2W_{\alpha} \alpha.
\end{equation}
In Fig. 2, this relation holds for $\alpha \in[0,1.6875]$ with $M=1$ and $M\in[0.7698,2]$ with $\alpha=1$. Here, $m(1,\alpha)=1(=M)$ while $m(M,1)$ is an increasing function of $M$.
\begin{figure}
\centering
\mbox{
(a)
\includegraphics[angle =0,scale=0.4]{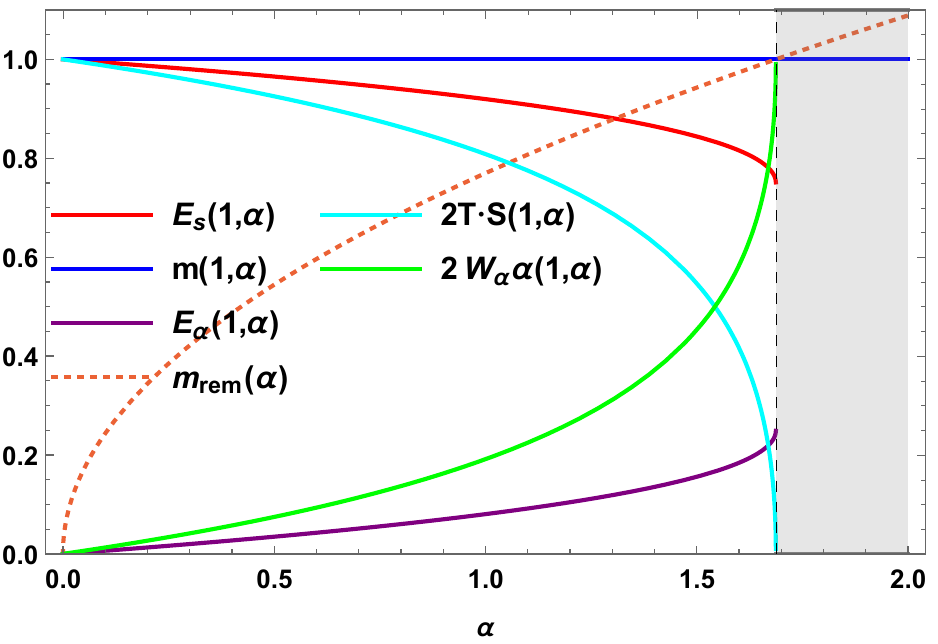}
(b)
\includegraphics[angle =0,scale=0.4]{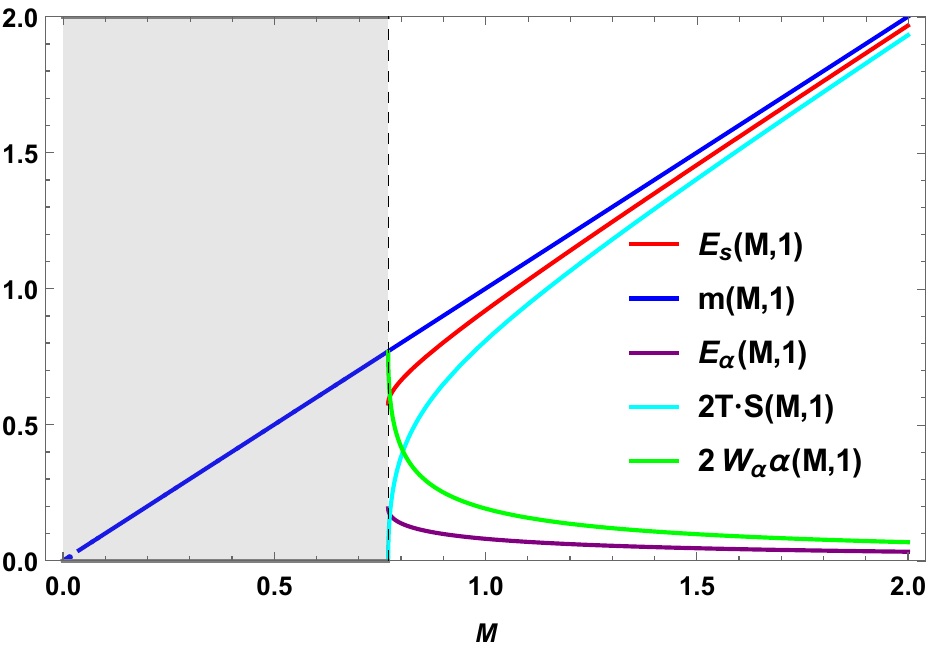}
}
\caption{Mass relation $E(=m)=E_S+E_\alpha$ and Smarr formula $m=2TS+2W_\alpha \alpha$. (a) Smarr formula and mass (internal energy) relation are  as functions of $\alpha\in[0,1.6875]$ with $M=1$. The remnant mass $m_{\rm rem}(\alpha)=0.7698 \sqrt{\alpha}$ is displayed. A shaded region of $\alpha\in(1.6875,2]$ is forbidden with $M=1$. (b)  Smarr formula and mass (internal energy) relation are defined  as functions of $M\in[0.7698,2]$ with $\alpha=1$. A shaded region of $M\in[0,0.7698)$ is forbidden with  $\alpha=1$.   }
\label{plot_modes}
\end{figure}
At this stage, we stress that the quantum parameter $\alpha$ is regarded as an important thermodynamic variable~\cite{Hajian:2023bhq}.
The remnant mass $m_{\rm rem}(\alpha)=0.7698 \sqrt{\alpha}$ is displayed in Fig. 2(a) for showing that  the APS dust ball has the bouncing property if $m<m_{\rm rem}(\alpha)$.

Let us study the Smarr formula.
We demonstrate that the black hole mass can be interpreted as the internal energy of the system corresponding to the sum of the surface energy  and the quantum energy  of the black hole.
Following the analysis in~\cite{Smarr:1972kt}, and since $ dm$ is a perfect differential in \eqref{first-law}, one is free to choose any convenient path of integration in the $(S,\alpha)$ space.
By choosing the path in an appropriate way, one is able to define two energy components for the solution \eqref{ansatz}.
The first one is the surface energy $E_s$  given by
\begin{equation}
	\label{eq:Es}
	E_s(S,\alpha)=\int_0^S T(\tilde{S},0) d \tilde{S}=\frac{\sqrt{S}}{2\sqrt{\pi}}\to E_s(M,\alpha)=\frac{r_+(M,\alpha)}{2},
\end{equation}
and is common even for Schwarzschild, Kerr, and Kerr-Newman black holes. On the other hand,  the second one is the quantum energy $E_\alpha$ related to the quantum parameter $\alpha$ and takes the form
\begin{eqnarray}
	\label{eq:Eq}
&&	E_\alpha(S,\alpha)=\int_0^\alpha W_\alpha(S,\tilde{\alpha}) d \tilde{\alpha}=m(S,\alpha)-\frac{\sqrt{S}}{2\sqrt{\pi}} \label{eq:Eq} \\
&& \to E_\alpha(M,\alpha)=\frac{r_+^3(M,\alpha)-r_+^2(M,\alpha)\sqrt{r_+^2(M,\alpha)-\alpha}}{\alpha}-\frac{r_+(M,\alpha)}{2}.\label{eq:EqS}
\end{eqnarray}
\begin{figure}
\centering
\mbox{
(a)
\includegraphics[angle =0,scale=0.4]{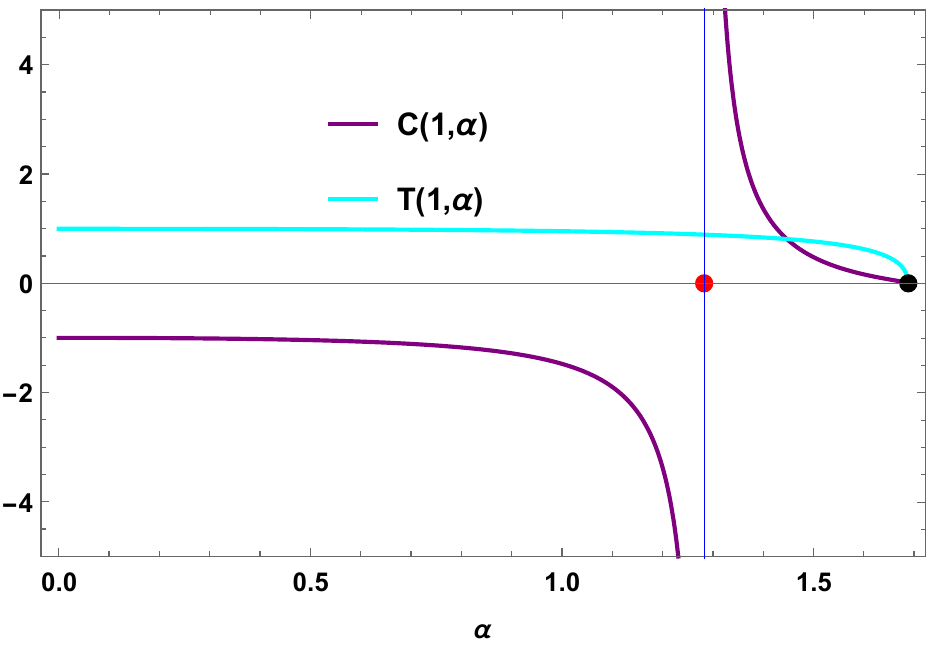}
(b)
\includegraphics[angle =0,scale=0.4]{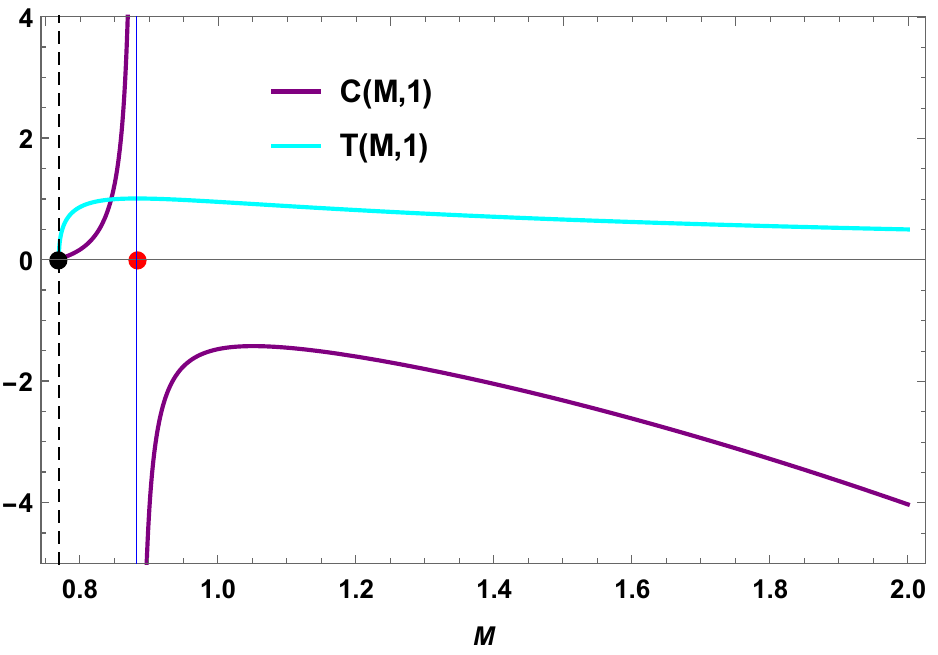}
}
\caption{ Heat capacity $C(1,\alpha)/|C_S(1,0)|$ with $|C_S(1,0)|=25.13$   (a) Heat capacity blows up at Davies point ($\alpha_D=1.2835,$ red dot) which is  the critical onset parameter ($\alpha_c$) for spontaneous scalarization.
The heat capacity and temperature $T(1,\alpha)/0.04$  are zero at the extremal point ($\alpha_e=1.6875,~\bullet$). (b) Heat capacity blows up at Davies point ($M_D=0.8827,$ red dot) where the temperature $T(M,1)$ has the maximum. This point coincides with the critical onset mass ($M_c$). The heat capacity and temperature are zero at the remnant point ($M_{\rm rem}=0.7698,~\bullet$). }
\label{plot_modes}
\end{figure}
Observing from Fig. 2, $m=E_s+E_\alpha$ indicates  that the mass of the black hole plays the role of the internal energy of the system.
This type of thermodynamic decomposition is also typical in GR black hole solutions, but the rotational energy and/or the electromagnetic energy appear~\cite{Smarr:1972kt}.
If one uses $r_{\rm +a}(M,\alpha)= 2M -\frac{\alpha}{8M}-\frac{3\alpha^2}{128}$ for $M\gg\alpha$, its energies are expressed in terms of $M$ and $\alpha$ as
\begin{equation}
E_s\simeq M-\frac{\alpha}{16M}-\frac{3\alpha^2}{256M^3},\quad E_\alpha\simeq \frac{\alpha}{16M}+\frac{3\alpha^2}{256M^3},\quad E\simeq M,
\end{equation}
which shows its proportions clearly: $E_s$ is major, while $E_\alpha$ is minor. One observes from Fig. 2(a) that for minors, $m_{\rm rem}(\alpha)>E_\alpha(1,\alpha),~2W_\alpha \alpha(1,\alpha)$  while  for majors, $m_{\rm rem}(\alpha)> 2T(1,\alpha)\cdot S(1,\alpha)$ for $\alpha>1.056$ and   $m_{\rm rem}(\alpha)>E_s(1,\alpha)$ for $\alpha>1.31$.

We observe from Fig. 3 that the heat capacity $C(1,\alpha)/|C_S(1,0)|$ blows up at  Davies point ($\alpha_D=1.2835$, red dot) which is exactly the critical onset parameter for spontaneous scalarization.
Also, the  heat capacity  $C(M,1)/|C_S(1,0)|$ blows up at Davies point ($M_D=0.8827$, red dot) where the temperature $T(M,1)$ takes  the maximum value. This point coincides with the critical onset mass ($M_c$). They are zero at extremal point ($\alpha_e=1.6875,~\bullet$) and remnant point ($M_{\rm rem}=0.7698,\bullet$).
Finally, the Gibbs free energy $G(1,\alpha)=1/2$ is a constant function of $\alpha$ and  $G(M,1)$ is an increasing function of $M$, showing no a particular feature.

\section{Shadow radius analysis}
We analyzed thermodynamics of qOS-black hole to find peculiar points and allowed regions for mass $M$ and quantum parameter $\alpha$, whose peculiar points are extremal point, Davies point, and remnant point.
In this section, we need to analyze shadow radii by computing the  photon radius and critical impact parameter to extend the allowed regions of $M\in[0.7698,$ ] and $\alpha\in[0,1.6875]$. The lower bound of $M$ and the upper bound of $\alpha$ can be extended to accommodate its NS (naked singularity)-branches by strong gravitational lenzing.
We will  compare shadow radii with the recent EHT observation.

Requiring  the photon sphere, one finds two conditions with a geodesic potential $\tilde{V}(r)=g(r)/r^2$
\begin{equation} \label{cond-LR}
\tilde{V}(r=L)=\frac{1}{2b^2}, \quad \tilde{V}'(r=L)=0,
\end{equation}
where $b$ is  the critical impact parameter and $L$ represents the radius  of unstable photon sphere.
 Eq.(\ref{cond-LR}) implies   two relations
 \begin{equation}
 L^2=g(L)b^2,\quad 2g(L)-Lg'(L)=0.
 \end{equation}
Here, the photon sphere radius and its critical impact parameter  are given by
\begin{eqnarray}
L(M,\alpha), \quad b(M,\alpha),\label{CI}
\end{eqnarray}
whose explicit forms are too complicated to write down here. It is important to  note that its $\alpha\in[1.6875,2.8477]$-NS and $M\in[0.6,0.7698]$-NS branches are arisen from the extensions of the photon sphere $L(M,\alpha)$.
At this stage, we wish to point out  that shadow analysis was performed for the qOS-black hole by making use of the approximate photon sphere $L_{\rm a}$ and critical impact parameter $b_{\rm a}$~\cite{Ye:2023qks}
\begin{equation}
L_{\rm a}=3M-\frac{\alpha}{9M},\quad b_{\rm a}=3\sqrt{3}M-\frac{\alpha}{6\sqrt{3}M}
\end{equation}
which can be obtained from $L(M,\alpha)$ and $b(M,\alpha)$ when using  the condition of  $M\gg\alpha$. In this case, one could not accommodate its NS-branches.

\begin{figure}
   \centering
   \mbox{
   (a)
  \includegraphics[width=0.4\textwidth]{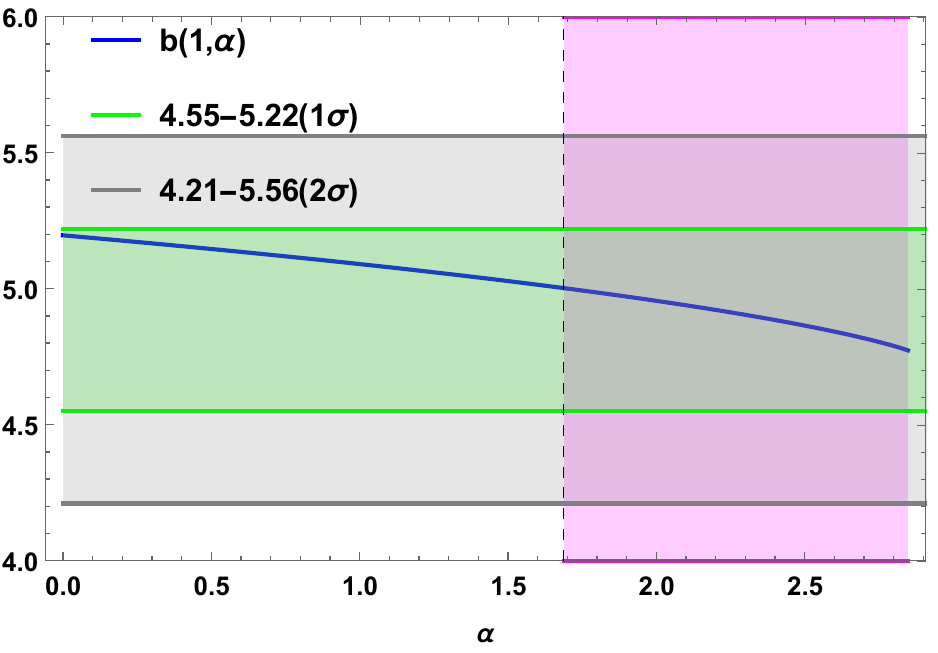}
 (b)
    \includegraphics[width=0.4\textwidth]{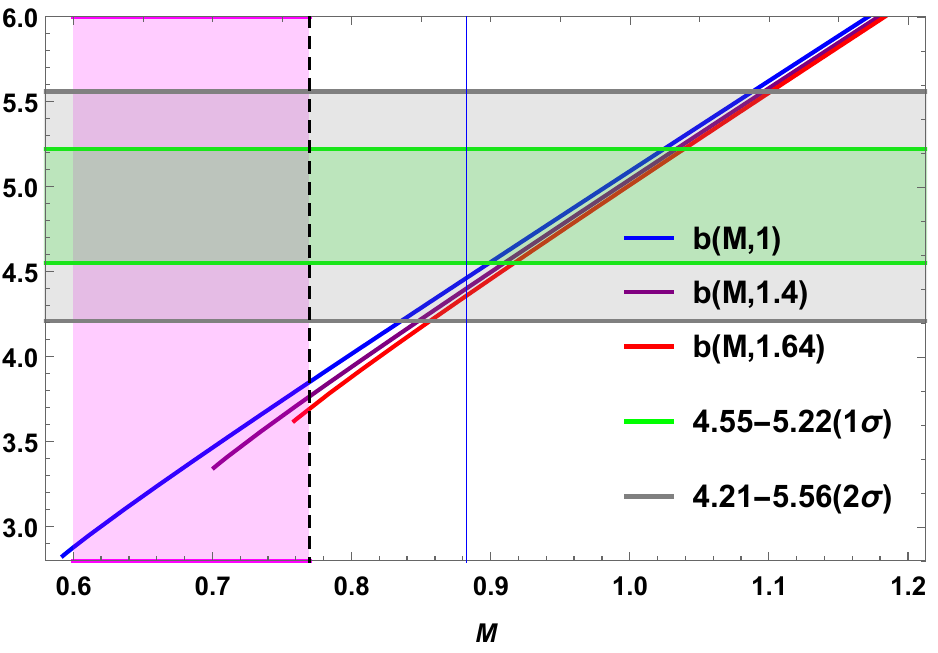}}
\caption{(a) Critical impact parameter  $b(M=1,\alpha)$ as a function of $\alpha\in[0,2.8477]$.  There is one shaded region $\alpha\in$[1.6875,2.8477] denoted by $\alpha$-NS region. A dashed line is at $\alpha_e=1.6875$ (extremal point).
 Here, we introduce $1\sigma$ and $2\sigma$ ranges from EHT observation.
 (b) Three critical impact parameters  $b(M,1),b(M,1.4),b(M,1.64)$ as functions of $M\in[0.6,1.2]$.  There is one shaded region $M\in$[0.6,0.7698] denoted by $M$-NS region.  A dashed line is at $M_{\rm rem}=0.7698$.  }
\label{cip_modes}
\end{figure}

Considering  the EHT observation (Keck- and VLTI-based estimates for SgrA$*$~\cite{EventHorizonTelescope:2022wkp,EventHorizonTelescope:2022wok,EventHorizonTelescope:2022xqj}), the  $1\sigma$ constraint on the shadow radius $r_{\rm sh}=b_i$ indicates ~\cite{Vagnozzi:2022moj}
\begin{equation}
4.55\lesssim r_{\rm sh} \lesssim 5.22  \label{KV1}
\end{equation}
and the  $2\sigma$ constraint shows
\begin{equation}
4.21 \lesssim r_{\rm sh} \lesssim 5.56. \label{KV2}
\end{equation}
Let us see  Fig. 4(a) for an explicit picture to test  with the EHT observation.
There is no constraints of the upper limit on its parameter  $\alpha$. This means that  within $1\sigma$, the qOS-branch and its $\alpha$-NS branch  including an extremal point ($\alpha_e=1.6875$) are consistent with the EHT observation~\cite{Vagnozzi:2022moj}.
From  Fig. 4(b), we find that the qOS-branch has two constraints on $M$ for $\alpha=1$: $0.8989\lesssim M \lesssim 1.024 (1\sigma)$ and  $0.8358\lesssim M \lesssim 1.088 (2\sigma)$.
Its $M$-NS branch is located beyond $2\sigma$.
 This is a new constraint on mass $M$ found from the qOS-black hole. For $\alpha=$ 1.4 and 1.64, one finds similar constraints on $M$.

\section{$\alpha-M$ induced spontaneous scalarizations}
Recently, it was shown that  the GB$^-$scalarization could  occur  when the
quantum  parameter satisfies $\alpha>\alpha_c=1.2835M^2$ in the EGBS theory with the unknown qOS action (\ref{Action1})~\cite{Chen:2025wze}.
In this section, we wish to revisit  the GB$^-$scalarization by computing critical onset mass and quantum parameter.
For this purpose, we need the scalar linearized equations which describe the scalar perturbation $\delta \phi$ propagating around (\ref{ansatz}).  It is derived by linearizing  Eq.(\ref{s-equa}) after choosing $f(\phi)=2\phi^2$ as
\begin{eqnarray}
  \left(\bar{\square}+ \lambda \bar{{\cal R}}^2_{\rm GB}\right)\delta \phi&=& 0. \label{l-eq2}
\end{eqnarray}
Here the overbar  denotes computation based on the qOS-black hole background (\ref{ansatz}).
Importantly,  we note that ``$-\lambda \bar{{\cal R}}^2_{\rm GB}$" plays a role of
 an effective mass $\tilde{m}^2_{\rm eff}$  for $\delta \phi$.
Introducing a tortoise coordinate defined by $dr_*=dr/g(r)$ and considering
\begin{equation}
\delta\phi(t,r_*,\theta,\varphi)=\sum_m\sum^\infty_{l=|m|}\frac{\psi_{lm}(t,r_*)}{r}Y_{lm}(\theta,\varphi),
\end{equation}
Eq.(\ref{l-eq2}) reduces  to the Klein-Gordon equation
\begin{equation} \label{mode-d}
\frac{\partial^2\psi_{lm}(t,r_*)}{\partial r_*^2} -\frac{\partial^2\psi_{lm}(t,r_*)}{\partial t^2}=V(r)\psi_{lm}(t,r_*),
\end{equation}
where the potential $V(r)$ is given by
\begin{equation} \label{pot-c}
V(r)=g(r)\Big[\frac{2M}{r^3}-\frac{4\alpha M^2}{r^6}+\frac{l(l+1)}{r^2}+\tilde{m}^2_{\rm eff}\Big]
\end{equation}
with its effective mass term
\begin{equation}
\tilde{m}^2_{\rm eff}=-\frac{48\lambda M^2}{r^{6}}\Big[\frac{3\alpha^2M^2}{r^6}-\frac{5\alpha M}{r^3} +1\Big].
\end{equation}
Eq.(\ref{mode-d}) is regarded as the (1+1)-dimensional mode decoupled equation  for $\psi_{lm}(t,r_*)$  because of a spherically symmetric background Eq.(\ref{ansatz}).
For $\lambda>0$ and $s$-mode ($l=0)$ with $\psi_{lm}(t,r_*)\sim u(r_*)e^{-i\omega t}$, one found  GB$^+$ scalarization of Schwarzschild black hole for large $\lambda$ and $\alpha=0$~\cite{Antoniou:2017acq,Doneva:2017bvd,Silva:2017uqg}.
For $\lambda<0$, one obtained spin-induced (GB$^-$) scalarization for rotating black holes with rotation parameter $a$~\cite{Dima:2020yac,Herdeiro:2020wei,Berti:2020kgk,Lai:2022spn,Lai:2022ppn}. In this study, we consider the $\lambda<0$ case with $l=0$ and $\alpha\not=0$. In this case, one has to find the  critical onset parameter $\alpha_c$ which determines the lower bound ($\alpha>\alpha_c$) for the onset spontaneous scalarization  by using the Hod's approach~\cite{Hod:2020jjy}. The onset of spontaneous scalarization  arises from  the potential.

To get the critical onset parameters, we consider the potential term only
\begin{equation}
V(r)\psi_{lm}(t,r_*)=0.
\end{equation}
The onset of spontaneous scalarization is related closely to  an effective  binding potential well in the near-horizon whose two turning points of $r_{\rm in}$ and $r_{\rm out}$ are classified by the relation of $r_{\rm out}\ge r_{\rm in}=r_+$.  A critical black hole with $\alpha=\alpha_c$  denotes the boundary between qOS-black hole and  scalarized qOS-black hole existing  in the limit of $\lambda \to -\infty$.
It is characterized by  the presence of a degenerate  binding potential well whose two turning points
merge at the outer horizon ($r_{\rm out}= r_{\rm in}=r_+$) as
\begin{eqnarray}
 \tilde{m}^2_{\rm eff}\psi_{lm}(t,r_*)=0, \quad {\rm for} \quad \alpha= \alpha_c
\end{eqnarray}
in the limit of $\lambda  \to -\infty$.
The critical onset (quantum)  parameter $\alpha_c$ is determined by the resonance condition because of $\psi_{lm}(t,r_*)\not=0$
\begin{eqnarray}
3\tilde{\alpha}_c^2-5\tilde{\alpha}_c+1 =0\label{res-con}
\end{eqnarray}
with $\tilde{\alpha}_c=\alpha_c M/r_+^3$.
Here we choose the small root
\begin{eqnarray}\label{critac}
\tilde{\alpha}_c=0.2324
\end{eqnarray}
for the critical onset parameter.
Solving $\tilde{\alpha}_c=\alpha_c/r_+^3(1,\alpha_c)=0.2324$ with $M=1$,  one finds
the critical onset  parameter for spontaneous scalarization being  the same  found in~\cite{Chen:2025wze} as
\begin{eqnarray}
\alpha_c=1.2835
\end{eqnarray}
which is consistent with the Davies point ($\alpha_D$).  Here, we check numerically that $\alpha_c(M,\alpha)=\alpha_D(M,\alpha)$ for $0.0128(M=0.1)\le \alpha \le 1.6875(M=1.14663)$ by solving  Eq.(\ref{res-con}) and $DC(M,\alpha)=0$ simultaneously.
This implies that the qOS-black holes with $\alpha <\alpha_c$ could not develop the tachyonic instability and could not have  scalarized qOS-black holes.
However, the other root of  $\tilde{\alpha}=1.4343$ to  Eq.~(\ref{res-con}) provides no physical solution for $\alpha_c$.

We note that  the Davies (critical) point is identified by the singular behavior of heat capacity at $\alpha=\alpha_D$ differing  from the extremal point ($\alpha_e=1.6875$).
 A second order phase transition occurs at this critical point
 and this phenomenon is generic for any charged or rotating black holes.   The nature of this
 critical phenomenon is not entirely clear except it has been known that the heat capacity
 changes its sign abruptly.

 Similarly, using the Hod's approach,  we find the critical onset mass by choosing the small root
 \begin{equation}
 M_c=0.8827
 \end{equation}
 which is the same as the Davies point of $M_D=0.8827$ found in the heat capacity $C(M,1)$.
  Here, one checks numerically that $M_c(M,\alpha)=M_D(M,\alpha)$ for $0.1(\alpha=0.0128)\le M \le 1.14663(\alpha=1.6875)$  by solving  Eq.(\ref{res-con}) and $DC(M,\alpha)=0$ simultaneously.
 This means that the qOS-black holes with $M>M_c$ could not develop the tachyonic instability and could not have  scalarized qOS-black holes.  The other root indicates no physical solution for $M_c$.
 Our model provides  the first example which shows that two Davies points ($\alpha_D,M_D$) are equal to two critical onset parameters ($\alpha_c,M_c$) for the $\alpha-M$ induced  scalarization.
Therefore, it is necessary to perform the GB$^-$ scalarization of  mass $M$ for qOS-black hole.
\begin{figure*}[t!]
   \centering
    \mbox{
   (a)
  \includegraphics[width=0.4\textwidth]{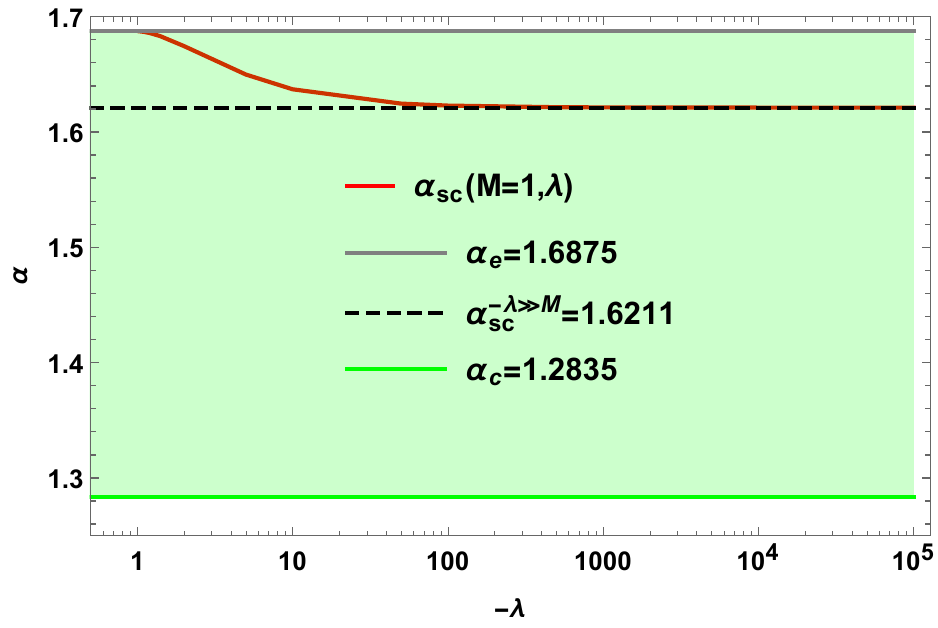}
 (b)
    \includegraphics[width=0.4\textwidth]{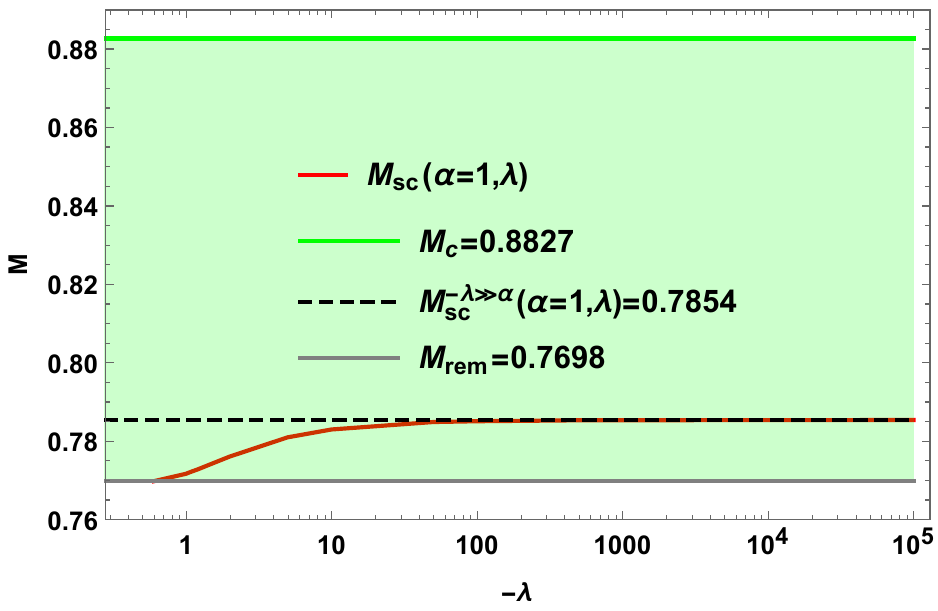}}
\caption{(a) Graph for sufficient condition  $\alpha_{sc}(M=1,\lambda)$ with the critical onset parameter $\alpha=\alpha_c$ as the lower bound.  A dashed line denotes the sufficient condition $\alpha=\alpha_{sc}^{-\lambda\gg M}$ for a large $-\lambda$. A top line denotes the extremal point at $\alpha=\alpha_e$ as the upper bound.
(b) Graph for $M_{sc}(\alpha=1,\lambda)$ with  the critical onset mass $M=M_c$ as the upper bound  and the sufficient condition $M=M_{sc}^{-\lambda\gg \alpha}$ for a large $-\lambda$.  A bottom line represents the remnant point at $M=M_{\rm rem}$ as the lower bound.}
\end{figure*}

To find the sufficient condition for the tachyonic instability, we use the  condition for instability suggested by Ref.\cite{Dotti:2004sh}
\begin{equation}
\int^\infty_{r_+(M,\alpha)}\Big[\frac{V(r,M,\alpha)}{g(r)}\Big] dr <0. \label{sc-ta}
\end{equation}
For large $-\lambda \gg M,\alpha$ where $M,\alpha$ are $\mathcal{O}(1)$,  this condition reduces to
\begin{equation}
\int^\infty_{r_+(M,\alpha)}\tilde{m}^2_{\rm eff} dr <0,
\end{equation}
which leads to
\begin{equation}
I(M,\alpha)=\frac{6M^2[120M^2 \alpha^2-275 M \alpha r_+^3(M,\alpha)+88r_+^6(M,\alpha)]}{55r_+^{11}(M,\alpha)}<0.
\end{equation}
This inequality  is not easy to solve for $\alpha$ because of the complexity of $r_+(M,\alpha)$.
Instead, we solve  $I(M,\alpha)=0$ to find two sufficient conditions  ($\alpha_{sc}^{-\lambda\gg M}=1.6212$ for $M=1,~M_{sc}^{-\lambda\gg  \alpha}=0.7854$ for $\alpha=1$) for large $-\lambda$. As is shown in Fig. 5, the allowed regions for tachyonic instability
are depicted  as $\alpha_c(=1.2835) <\alpha<\alpha_e(=1.6875) $ (green shaded region) for $ M=1$  and $M_{\rm rem}(=0.7698)<M<M_c(=0.8827) $ (green shaded region) for $\alpha=1$. The lower region of $0<\alpha<\alpha_c$ and the upper region of $M>M_{sc}$ are not suitable  for spontaneous scalarization.

To obtain $\alpha_{sc}(M=1,\lambda)$ and $M_{sc}(\alpha=1,\lambda)$, we have to find the condition of Eq.(\ref{sc-ta}) for a given $\lambda$ numerically.
We find that  $\alpha_{sc}(M=1,\lambda)$ is a decreasing function, connecting between  $\alpha_e$ and  $\alpha_{sc}^{-\lambda\gg M,\alpha}$, while $M_{sc}(\alpha=1,\lambda)$ is an increasing function, connecting between $M_{\rm rem}$ and
$M_{sc}^{-\lambda\gg  M,\alpha}$.

Finally, we would like to mention that to derive $\alpha_{th}(M=1,\lambda)$ connecting between $\alpha_c$ and $\alpha_e$ and $M_{th}(\alpha=1,\lambda)$ connecting between $M_{\rm rem}$ and $M_c$, one has to solve the Klein-Gordon equation (\ref{mode-d}) numerically with initial Gaussian wave packet~\cite{Chen:2025wze}. In this case, the fourth-order Runge-Kutta method plays a primary role in computing the time-domain profile of the scalar and the finite difference approach is introduced to validate the results. This is because the full onset for  GB$^-$ scalarization is a complicated phenomena~\cite{Lai:2022spn,Lai:2022ppn}.

\section{$\alpha-M$ induced spontaneous scalarizations with $\mathcal{L}_{\rm NED}$ }
In this section, we wish to perform $\alpha-M$ induced spontaneous scalarizations by choosing  $\mathcal{L}_{\rm NED}$ in Eq.(\ref{NED}).
Here, its metric function is given by
\begin{equation}
\tilde{g}(r)=1-\frac{2M}{r}+\frac{\alpha P^2}{r^4},
\end{equation}
where $P$ denotes the magnetic charge. From $\tilde{g}(r)=0$, we define $r_\pm(M,\alpha,P)$ for the outer/inner horizons firstly.  Its thermodynamic quantities are found  differently  as
\begin{eqnarray}
\tilde{m}(M,\alpha,P)&=&\frac{\alpha P^2+r_+^4(M,\alpha,P)}{r_+^3(M,\alpha,P)},  \nonumber \\
\tilde{T}(M,\alpha,P)&=&\frac{-3\alpha P^2+r_+^4(M,\alpha,P)}{4\pi r_+^5(M,\alpha,P)},  \label{N-ther} \\
\tilde{C}(M,\alpha,P)&\equiv& \frac{\tilde{N}C(M,\alpha,P)}{\tilde{D}C(M,\alpha,P)}=\frac{2\pi r_+^2(M,\alpha,P)[ 3\alpha P^2-r_+^4(M,\alpha,P)]}{ r_+^4(M,\alpha,P)-15 \alpha P^2},  \nonumber \\
\end{eqnarray}
where one checks that $\tilde{T}(M,\alpha,P)=\tilde{T}_\kappa(M,\alpha,P)$.
The first law and Smarr formula are defined as
\begin{equation}
d\tilde{m}=\tilde{T}d\tilde{S}+\tilde{W}_\alpha d\alpha,\quad \tilde{m}=2\tilde{T}\tilde{S}+4\tilde{W}_\alpha \alpha.
\end{equation}
In this case, the scalar potential takes the form
\begin{equation} \label{pot-c}
\tilde{V}(r)=\tilde{g}(r)\Big[\frac{2M}{r^3}-\frac{4\alpha P^2}{r^6}+\frac{l(l+1)}{r^2}+\hat{m}^2_{\rm eff}\Big]
\end{equation}
with its effective mass term
\begin{equation}
\hat{m}^2_{\rm eff}=-\frac{48\lambda M^2}{r^{6}}\Big[\frac{3\alpha^2P^2}{r^6}-\frac{5\alpha P}{r^3} +1\Big].
\end{equation}
Here, we find two critical onset parameters with $M=1,P=0.6$ and $\alpha=1,P=0.6$
\begin{equation}
\alpha_c=2.4948,~\quad M_c=0.7556
\end{equation}
which are not the same as  the Davies points of ($\alpha_D=2.2888,~M_D=0.8130)$ obtained from $\tilde{D}C(1,\alpha,0.6)=0$ and $\tilde{D}C(M,\alpha=1,0.6)=0$.
This shows clearly that  the metric functions  $g(r)$ for the unknown  $\mathcal{L}_{\rm qOS}$  is  quite different from $\tilde{g}(r)$ for $\mathcal{L}_{\rm NED}$.
\begin{figure*}[t!]
   \centering
    \mbox{
   (a)
  \includegraphics[width=0.4\textwidth]{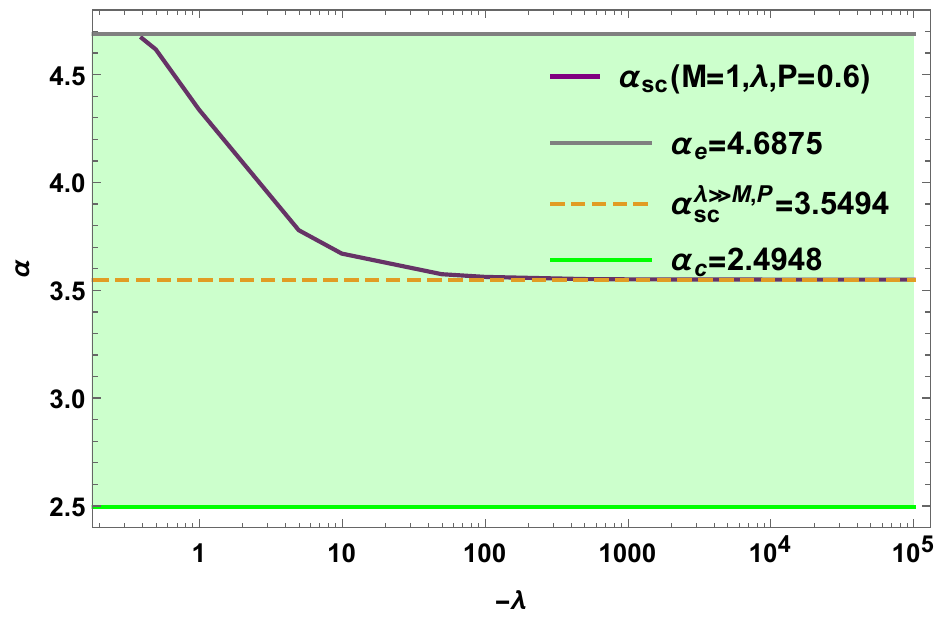}
 (b)
    \includegraphics[width=0.4\textwidth]{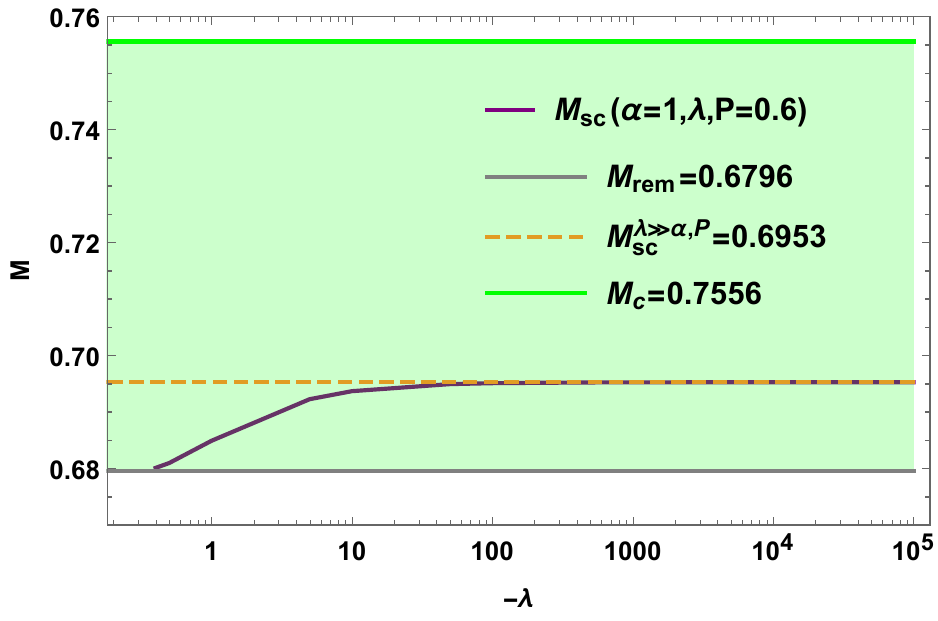}}
\caption{(a) Sufficient condition of $\alpha_{sc}(M=1,\lambda,P=0.6)$ with the critical onset parameter $\alpha=\alpha_c$ as the lower bound.  A dashed line denotes the sufficient condition $\alpha=\alpha_{sc}^{-\lambda\gg M,P}=3.5494$ for a large $-\lambda$. A top line denotes the extremal point at $\alpha=\alpha_e=4.6875$ as the upper bound.
(b) Graph for $M_{sc}(\alpha=1,\lambda,P=0.6)$ with  the critical onset mass $M=M_c$ as the upper bound  and the sufficient condition $M=M_{sc}^{-\lambda\gg  \alpha,P}=0.6953$ for a large $-\lambda$.  A bottom line represents the remnant point at $M=M_{\rm rem}=0.6796$ as the lower bound.}
\end{figure*}
To find  $\alpha_{sc}(M=1,\lambda,P=0.6)$ and $M_{sc}(\alpha=1,\lambda,P=0.6)$, we analyze  the condition of
\begin{equation}
\int^\infty_{r_+(M,\alpha,P)}\Big[\frac{\tilde{V}(r,M,\alpha,P)}{\tilde{g}(r)}\Big] dr <0 \label{sc-pta}
\end{equation}
for a given $\lambda$ numerically.
As is depicted in Fig. 6, we find that  $\alpha_{sc}(M=1,\lambda,P=0.6)$ is a decreasing function, connecting between  $\alpha_e$ and  $\alpha_{sc}^{-\lambda\gg M,P}$, while $M_{sc}(\alpha=1,\lambda,P=0.6)$ is an increasing function, connecting between $M_{\rm rem}$ and
$M_{sc}^{-\lambda\gg \alpha,P}$. We note that Fig. 6 is similar to Fig. 5.

\section{Discussions}

First of all, we obtained quantum number ($\alpha$)-mass ($M$) induced spontaneous  scalarization of qOS-black hole in the Einstein-Gauss-Bonnet-scalar theory with the unknown qOS action.
This study  is the first example to show double ($\alpha,M$) induced scalarizations  for a spherically symmetric qOS black hole, comparing to spin-charge induced scalarization for Kerr-Newman black holes in the Einstein-Maxwell-scalar theory~\cite{Lai:2022ppn}.

We have performed  thermodynamic aspects   by correcting thermodynamic quantities, checking  the first law thermodynamics and the Smarr formula, and studying a phase transition at Davies point appeared in heat capacity.

Importantly, we found a close connection between thermodynamics of bald black hole and spontaneous scalarization.
It was shown that two Davies points of heat capacity are identified with two critical onset mass and quantum parameter for spontaneous scalarization in  the EGBS theory with the unknown qOS action.
We point out that  such connections are not found from the Einstein-Gauss-Bonnet-scalar theory with the nonlinear electrodynamics action proposed by~\cite{Mazharimousavi:2025lld}.
As far as we know, there is no model to show  such a connection between thermodynamics of bald black hole and spontaneous scalarization.

Hence, it is necessary to perform the full scalarization of quantum number ($\alpha$) and  mass ($M$)  to obtain  scalarized  qOS-black holes  by solving full equations (\ref{equa1}) and (\ref{s-equa}).
For this purpose, one  may allow to  use the nonlinear electrodynamics action Eq.(\ref{NED}) for completeness, instead of the unknown qOS action $\mathcal{L}_{\rm qOS}$, because its onset of spontaneous scalarization was performed here.

\section{Acknowledgments}

Y.S.M. is supported by the National Research Foundation of Korea (NRF) grant
 funded by the Korea government (MSIT) (RS-2022-NR069013).

\end{document}